\documentclass[10pt,twocolumn,letterpaper]{article}

\usepackage[pagenumbers]{cvpr} 

\usepackage{graphicx}
\usepackage{caption}
\usepackage{booktabs}
\usepackage{amssymb}
\usepackage{pifont}
\usepackage{amsmath}
\usepackage{makecell}
\usepackage{acro}

\newcommand{\compactparagraph}[1]{%
    \par\addvspace{0.2\baselineskip}%
    \noindent\textbf{#1}\nobreak\hspace{0.5em}\ignorespaces
}

\makeatletter

\newcommand{\Gaussian}{\ensuremath{G}\xspace}
\newcommand{\mean}{\ensuremath{\mu}\xspace}
\newcommand{\cov}{\ensuremath{\Sigma}\xspace}
\newcommand{\opacity}{\ensuremath{o}\xspace}
\newcommand{\compcolor}{\ensuremath{sh}\xspace}
\newcommand{\NumberOfGaussians}{\ensuremath{N}\xspace}
\newcommand{\parameters}{\ensuremath{\theta}\xspace}

\newcommand{\NumberOfFrames}{\ensuremath{K}\xspace}

\newcommand{\NumberOfCams}{\ensuremath{M}\xspace}

\newcommand{\CurrentTime}{\ensuremath{t}\xspace}

\newcommand{\BlockSize}{\ensuremath{B}\xspace}
\newcommand{\CurrentBlock}{\ensuremath{k}\xspace}
\newcommand{\Block}{\ensuremath{\mathcal{B}}\xspace}
\newcommand{\WorkingSet}{\ensuremath{\mathcal{K}}\xspace}

\newcommand{\TemporalCenter}{\ensuremath{g}\xspace}
\newcommand{\TemporalScale}{\ensuremath{u}\xspace}
\newcommand{\Velocity}{\ensuremath{\mathbf{v}}\xspace}
\newcommand{\Acceleration}{\ensuremath{\mathbf{a}}\xspace}
\newcommand{\Jerk}{\ensuremath{\mathbf{j}}\xspace}
\newcommand{\Snap}{\ensuremath{\mathbf{s}}\xspace}

\newcommand{\BlockCenter}{\ensuremath{\mathbf{c}}\xspace}
\newcommand{\BlockRadius}{\ensuremath{r}\xspace}
\newcommand{\MotionBound}{\ensuremath{\rho}\xspace}
\newcommand{\TemporalSupport}{\ensuremath{S}\xspace}
\newcommand{\OpacityThreshold}{\ensuremath{\epsilon}\xspace}

\DeclareAcronym{gs}{
  short = GS,
  long  = Gaussian Splatting
}

\DeclareAcronym{doom-gs}{
  short = DOOM-GS,
  long  = Dynamic Out-of-Core Optimization for Massive Gaussian Splatting
}

\DeclareAcronym{gps}{
  short = GPS,
  long  = Gaussian Point Splatting
}

\definecolor{cvprblue}{rgb}{0.21,0.49,0.74}
\usepackage[pagebackref,breaklinks,colorlinks,allcolors=cvprblue]{hyperref}

\title{DOOM-GS: Dynamic Out-of-core Optimization for Massive Gaussian Splatting}

\author{
Do\u{g}a Y{\i}lmaz$^{1}$ \quad
Jialin Zhu$^{2}$ \quad
Binglun Wang$^{1}$ \quad
He Wang$^{1}$\\
$^{1}$University College London \qquad
$^{2}$Baidu Research\\
{\small\ttfamily \{doga.yilmaz,binglun.wang,he\_wang\}@ucl.ac.uk}\\
{\small\ttfamily misaliet@outlook.com}
}

\begin{document}
\maketitle
\begin{abstract}
Traditional Gaussian Splatting (GS) methods use in-core training, i.e. assuming that GPU memory is large enough to accommodate all the primitives. 
This assumption breaks down when the primitives exceed the capacity of GPU memory, e.g. GS on large environments. 
Recent out-of-core training research starts to address this issue, but is limited to static scenes. 
Unlike static GS, dynamic GS primitives move in space and time, breaking common assumptions held in existing out-of-core research and impose new challenges in memory management and training. 
To this end, we propose Dynamic Out-of-core Optimization for Massive Gaussian Splatting (DOOM-GS), the first out-of-core optimization method designed specifically for massive-scale dynamic GS. 
DOOM-GS is equipped with novel memory management strategies that optimize caching for spatio-temporal primitives, and a fast learning method for large numbers of primitives. 
Due to the lack of suitable datasets for large-scale dynamic GS, we also contribute a new dataset for evaluation. 
Since there are no directly comparable methods, we adapt several closely related methods for comparison. 
Through extensive evaluation, we demonstrate the effectiveness of DOOM-GS which now makes it possible to reconstruct city-scale dynamic scenes with high quality on moderate hardware, serving as a foundational tool for diverse future research and applications.
\end{abstract}    
\section{Introduction}
\label{sec:intro}

\ac{gs} has become a foundational component of tools for graphics~\cite{kerbl20233d,yang2023gs4d,kong2025generative}, robotics~\cite{wu2025rl,shen2026lyra,jiang2026gsworld}, medical imaging~\cite{kleinbeck2025multi,yu2025mathbf,wang2026medgssr}, \etc. 
Given its universal importance, recent research has attempted to fundamentally improve \ac{gs}' capabilities with more flexible representations~\cite{zhu20253d,liu2025deformable,diolatzis2024n,song2025spline,wang2025freetimegs}, improved optimization processes~\cite{kheradmand20243d,rota2024revising,sandu2026temporally}, and incorporated data uncertainty~\cite{kim20244d,hanson2025pup,yilmaz2026graphixs}. 
Being orthogonal to all these attempts, a fundamental compromise in \ac{gs} is the number of primitives versus the difficulty of training. 
Generally, more primitives lead to higher quality but require more computation, larger memory consumption, and thus longer and brittle training. 
Furthermore, since most training methods require full GPU residency of all primitives, it puts a significant constraint on the number of primitives.

To address this, recent out-of-core methods have enabled \ac{gs} training on large static scenes by dynamically transferring primitives between GPU, CPU, and storage~\cite{zhong2026tidegs,zhao2026clm,windisch2026ALoDOfGaussians}. 
However, extending this capability to dynamic scenes is substantially more challenging, as motion and finite temporal support of components violate two key assumptions of existing methods. 
First, existing methods exploit static spatial locality, meaning that Gaussians that are close in 3D space are stored or transferred together and are expected to remain together throughout training~\cite{zhong2026tidegs,zhao2026clm,windisch2026ALoDOfGaussians}. 
In contrast, Gaussians in dynamic scenes move and remain active only over finite temporal intervals, causing static spatial groupings to lose locality over time and thereby violating this assumption. 
Second, the primitives visible from a single view may already exceed the GPU memory, breaking the assumption that GPU memory can accommodate primitives across views for resident blocks re-usage~\cite{zhong2026tidegs,zhao2026clm}. 
A new approach addressing these two challenges is needed.

To enable dynamic \ac{gs} on large scenes, we propose \ac{doom-gs}, which is the first dynamic out-of-core \ac{gs} training framework to the best of our knowledge. 
\ac{doom-gs} addresses the first challenge by maintaining compact space-time blocks, in contrast to spatial blocks in existing research~\cite{zhong2026tidegs,zhao2026clm,windisch2026ALoDOfGaussians}, via a series of bespoke operations. 
In addition, it employs overlapping-aware camera scheduling to reduce data transfers while preserving camera and temporal diversity for stable training. 
Second, to remove the per-view memory constraint of existing methods, we propose a fast learning approach by augmenting an inference-only method \ac{gps}~\cite{rijsdijk2026gps} into a learnable model, relaxing the constraint imposed by GPU on the number of primitives associated with a view.

Evaluation is also challenging because no existing method directly supports massive dynamic out-of-core training, and no public benchmark combines dynamic content, dense multi-view observations, and large-scale coverage. 
Therefore, we introduce DynamicMatrixCity, a new dataset based on MatrixCity~\cite{li2023matrixcity} containing dynamic objects. 
We also adapt the closest out-of-core methods for static scenes as baselines. 
Our experiments demonstrate that \ac{doom-gs} enables high-quality city-scale dynamic reconstruction under bounded GPU memory while improving temporal consistency and training efficiency. 
Our contributions are:
\begin{itemize}
    \item \ac{doom-gs}, the first dynamic out-of-core \ac{gs} training framework for massive dynamic scenes.
    \item A new data block management strategy and camera scheduling for efficient data transfer.
    \item A new learnable rendering process, allowing learning from views where the primitives associated with a single view are too large to fit in GPU memory.
    \item A new dataset DynamicMatrixCity, a city-scale multi-view dataset for out-of-core dynamic \ac{gs}.
\end{itemize}

\section{Related Work}
\label{sec:rel_work}

\compactparagraph{Dynamic Gaussian Splatting.} 
Most dynamics GS methods differ mainly in their representations of motion. 
3D Gaussians can be mapped to different timesteps~\cite{yang2024deformable,wu20244d,wang2026ode}.
Alternatively, spatio-temporal behaviors can be explicitly encoded into primitives. 
They can be directly optimized over space and time~\cite{yang2024real, duan20244d}, or parameterized by polynomial functions~\cite{li2024spacetime,song2025spline}. 
The dynamics of Gaussians can be limited to a short temporal support~\cite{wang2025freetimegs} to reduce redundancy. 
High-order dynamics can improve the modeling of complex motions~\cite{yilmaz2026graphixs}. 
Our work is orthogonal to this research and adopts the explicit space-time representation and focuses on scaling its optimization to massive dynamic scenes.

\compactparagraph{Out-of-Core Optimization.}
Unlike conventional \ac{gs} methods, recent out-of-core training methods can optimize large \ac{gs} scenes on moderate GPUs~\cite{zhao2026clm,zhong2026tidegs,windisch2026ALoDOfGaussians}. 
CLM-GS~\cite{zhao2026clm} offloads Gaussian parameters to CPU memory while overlapping data transfer with GPU computation.
TideGS~\cite{zhong2026tidegs} introduces an SSD--CPU--GPU hierarchy with asynchronous streaming and prefetching, while A LoD of Gaussians~\cite{windisch2026ALoDOfGaussians} combines out-of-core training with a level-of-detail representation. 
These methods partition the model into fixed-size blocks enabling efficient contiguous transfers and shifts the model-size limit from GPU memory to storage. 
However, they target static scenes and remain limited by storage traffic, poor spatial and temporal block locality, and gradient computation restricted to GPU-resident Gaussians. 
\ac{doom-gs} extends this paradigm to dynamic scenes while addressing these limitations.

\compactparagraph{Rendering Beyond GPU Memory.}
Despite advances in out-of-core training for \ac{gs}, existing methods still face two key challenges. 
They either rely on complex hierarchical level-of-detail structures~\cite{shen2025lod,cheng2026clod,windisch2026ALoDOfGaussians} or cannot process views where the visible Gaussians exceed GPU memory capacity. 
To this end, various rendering methods have been proposed to render scenes that contain more components than the GPU memory limit. 
StochasticSplats~\cite{kheradmand2025stochasticsplats} is one of the first works to propose sort-free rendering of scenes by leveraging stochastic rasterization and an unbiased Monte Carlo estimator of the rendering equation. 
Although this is not explicitly designed to address the GPU memory limit, it enables processing the alpha blending in parts by eliminating the need for sorting. 
Following this, \acl{gps}~\cite{rijsdijk2026gps} proposed a stochastic rendering method that samples pixel-sized opaque points from Gaussians and splats them independently, enabling scalable rendering of scenes beyond GPU memory limits. 
However, it is only for forward rendering and cannot be used for learning. 
We turn it into a learnable process, enabling learning from views where associated Gaussians are too big for GPU.

\compactparagraph{Large-Scale Datasets.}
Capturing large multi-view dynamic scenes for benchmarking \ac{gs} methods remains challenging. 
Existing benchmarks are generally either large-scale but static~\cite{li2023matrixcity} or dynamic but limited in scene size and camera coverage~\cite{pumarola2020d, li2022neural, broxton2020immersive}. 
Furthermore, acquiring large-scale real-world data requires dense arrays of synchronized and calibrated cameras over large areas, while raising substantial logistical, storage, and privacy concerns. 
Consequently, synthetic environments remain the most practical option for generating city-scale dynamic observations. 
MatrixCity~\cite{li2023matrixcity} provides full-city scenes at multiple scales, but its publicly released data contain only static content.

\section{Methodology}
\label{sec:methodology}

\subsection{Preliminaries}
\label{sec:preliminaries}
\compactparagraph{Component Dynamics.}
Dynamic \ac{gs} makes each Gaussian component time-dependent, allowing a single model to represent a dynamic scene. 
We follow~\cite{yilmaz2026graphixs}, where each component~$\Gaussian_i$ is parameterized by $\parameters_i$, comprising its mean~$\mean_i$, covariance~$\cov_i$, opacity~$\opacity_i$, color~$\compcolor_i$, appearance time~$g_i$, duration~$u_i$, and motion parameters $\{\mathbf{v}_i,\mathbf{a}_i,\mathbf{j}_i,\mathbf{s}_i\}$, representing velocity, acceleration, jerk, and snap. 
At normalized time $\CurrentTime\in[0,1]$, its mean and opacity are
\begin{align}
    &\mean_i(\CurrentTime) = \mean_i + f_i(\CurrentTime), \opacity_i(\CurrentTime) = \opacity_i
    \exp\!\left(-\frac{(\CurrentTime-g_i)^2}{2u_i^2}\right) \nonumber\\
    &\text{where } f_i(\CurrentTime) =
    \mathbf{v}_i\Delta t
    + \frac{\mathbf{a}_i}{2}\Delta t^2
    + \frac{\mathbf{j}_i}{6}\Delta t^3
    + \frac{\mathbf{s}_i}{24}\Delta t^4.
\label{eq:dynamic_gs}
\end{align}
$\Delta t$ is the time step size. $\TemporalCenter_i$ and $\TemporalScale_i$ jointly control the temporal change of opacity of each component. Covariance and color remain fixed in time. Finally, each frame is rendered from the moving and time-parameterized Gaussians. 

\compactparagraph{Point Splatting.}
Standard \ac{gs} renders Gaussians using depth-ordered alpha compositing. 
\ac{gps}~\cite{rijsdijk2026gps} instead represents each projected Gaussian with independently sampled opaque points and retains the nearest point at each subpixel through an atomic depth test.
We adopt this renderer because its order-independent formulation does not require all visible Gaussians to reside in GPU memory simultaneously for sorting.
Let $\opacity_i(q)$ denote the opacity of Gaussian $i$ at subpixel $q$, and let $\lambda_i(q)$ be the expected number of points it generates at that location.
For a Poisson process with rate $\lambda_i(q)$, the probability of generating no point is $e^{-\lambda_i(q)}$.
Matching this probability to the Gaussian transparency $1-\opacity_i(q)$ gives
\begin{equation}
e^{-\lambda_i(q)}
=
1-\opacity_i(q),
\qquad
\lambda_i(q)
=
-\ln\bigl(1-\opacity_i(q)\bigr).
\label{eq:preliminary_point_splatting_1}
\end{equation}
Thus, Gaussian $i$ generates at least one point at $q$ with probability $\opacity_i(q)$.
A point from Gaussian $i$ is visible only if no Gaussian in front of it generates a point at the same subpixel.
The corresponding transmittance is
\begin{equation}
T_i(q)
=
\prod_{j\text{ in front of }i}
\bigl(1-\opacity_j(q)\bigr).
\end{equation}
Gaussian $i$ therefore provides the nearest point with probability $\opacity_i(q)T_i(q)$, matching its weight in standard alpha compositing.
Consequently, the expected point-splatting image approximates the standard \ac{gs} image.

\subsection{DOOM-GS}
\begin{figure*}[t!]
    \centering
    \includegraphics[width=\textwidth]{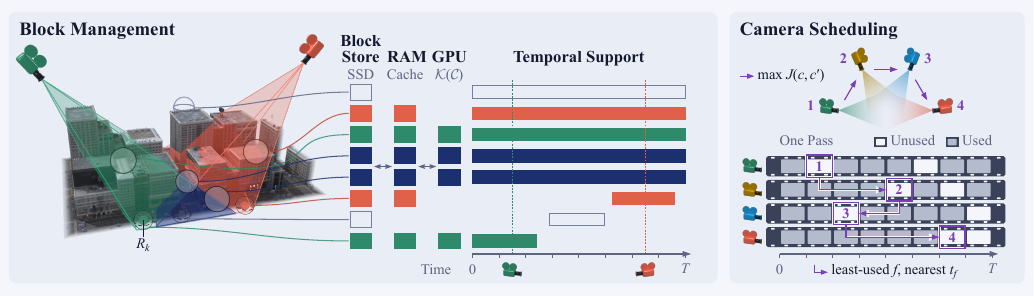}
    \caption{\textbf{\ac{doom-gs} overview.} \emph{Left:} Gaussians are stored on SSD in fixed-size blocks, with frequently accessed blocks cached in CPU memory. A block enters the dynamic working set $\WorkingSet(\mathcal{C})$ when its motion-aware bounding sphere of radius $R_k$ is spatially visible and the camera timestamp lies within its temporal support $\TemporalSupport_k$ as in Equation~\ref{eq:dynamic_working_set}. Green and orange blocks are selected by one camera, dark blue blocks by both, and hollow blocks by neither due to spatial or temporal culling. \emph{Right:} Each scheduling pass selects one view from every camera. The next unused camera maximizes $J(c,c')$, and its least-used frame is selected with ties resolved by proximity to $t_f$. 
    }
    \label{fig:doomgs}
\end{figure*}
\ac{doom-gs} trains massive dynamic \ac{gs} models on a single 24\,GB GPU using standard CPU memory and SSD storage.
As illustrated in Figure~\ref{fig:doomgs}, Gaussians are stored in fixed-size SSD blocks, frequently accessed blocks are cached in CPU memory, and only blocks visible to the current camera batch are loaded onto the GPU.
Moving out-of-core training from static to dynamic scenes raises three challenges. 
First, Gaussians move and exist only for finite time spans, so blocks based on purely spatial grouping quickly lose their locality. 
Next, the visible set of Gaussians changes with both the camera and time. 
Last, a single view of a large scene can already involve Gaussians exceeding the capacity of GPU memory. 
\ac{doom-gs} addresses these challenges with space-time-aware block organization, overlap-aware camera scheduling for block reuse, and chunked point splatting with gradient estimators.

\subsubsection{Block Storage for Dynamic Gaussians}
Similar to TideGS~\cite{zhong2026tidegs}, we store the $\NumberOfGaussians$ Gaussians in fixed-size blocks containing $\BlockSize=4096$ records.
This results in $K_{\mathrm{blk}}=\NumberOfGaussians/\BlockSize$ blocks.
We index the blocks by $k\in\{1,\ldots,K_{\mathrm{blk}}\}$ and denote the set of Gaussians stored in block $k$ by $\Block(k)$. 
All blocks are stored on SSD in an out-of-core structure that we refer to as the block store.
Each record stores 59 static parameters, comprising its 3D mean $\mean_i$, covariance $\cov_i$, opacity $\opacity_i$, and spherical-harmonic coefficients $\compcolor_i$.
An additional 14 parameters describe its dynamics through the temporal center $\TemporalCenter_i$, temporal scale $\TemporalScale_i$, and four 3D motion coefficients $\{\Velocity_i,\Acceleration_i,\Jerk_i,\Snap_i\}$.
Each block $\CurrentBlock$ is represented by a bounding sphere with center and radius $(\BlockCenter_k,\BlockRadius_k)$ that encloses the means of all Gaussians in that block. 
Given a batch of cameras $\mathcal{C}$, the system loads a block if its bounding sphere intersects the viewing frustum of at least one camera in the batch. 
However, this spatial criterion alone is insufficient for dynamic Gaussians whose means vary over time and whose contributions are limited to finite temporal intervals.
We therefore extend block-level culling to jointly account for spatial visibility and temporal activity.
Let $\OpacityThreshold$ denote the minimum opacity that we consider significant. For a Gaussian temporal profile in Equation~\ref{eq:dynamic_gs}, the opacity decreases with the distance from its temporal center. 
The factor $\kappa=\sqrt{-2\ln\OpacityThreshold}$ gives the distance, in units of temporal scale, at which the opacity reaches $\OpacityThreshold$. 
We therefore consider Gaussian $i$ active when $|\CurrentTime-\TemporalCenter_i| \leq \kappa\TemporalScale_i.$
For block $\CurrentBlock$, we define a common temporal margin
\begin{equation}
\Delta_k = \max_{i\in\Block(k)}\kappa\TemporalScale_i.
\end{equation}
The temporal support of the block is then
\begin{equation}
\TemporalSupport_k =
\left[
\min_{i\in\Block(k)}\TemporalCenter_i-\Delta_k,\,
\max_{i\in\Block(k)}\TemporalCenter_i+\Delta_k
\right].
\end{equation}
This interval is the union of the active periods of all Gaussians in block $k$. We additionally account for the motion of each Gaussian within its temporal support and denote the temporal support radius of Gaussian $i$ by $\delta_i = \kappa\TemporalScale_i$. We use $\delta_i$ as the maximum distance in time from the temporal center $g_i$ within which Gaussian $i$ is considered active. 
Using $\delta_i$, we can derive the maximum displacement in space by
\begin{equation}
\MotionBound_i = \|\Velocity_i\|\delta_i
+\frac{\|\Acceleration_i\|}{2}\delta_i^2
+\frac{\|\Jerk_i\|}{6}\delta_i^3
+\frac{\|\Snap_i\|}{24}\delta_i^4.
\end{equation} 
where $\MotionBound_i$ is the maximum displacement of Gaussian $i$ within its temporal support.
We extend the bounding sphere of each block by the largest $\MotionBound_i$ among its Gaussians:
\begin{equation}
R_k = \BlockRadius_k + \max_{i\in\Block(k)}\MotionBound_i.
\end{equation}
For a camera batch $\mathcal{C}$, the dynamic working set is
\begin{equation}
\WorkingSet(\mathcal{C})
=
\left\{
k \;\middle|\;
\exists c\in\mathcal{C}:
t_c\in\TemporalSupport_k
\land
\operatorname{visible}(\BlockCenter_k,R_k,c)
\right\},
\label{eq:dynamic_working_set}
\end{equation}
where $t_c$ is the timestamp of camera $c$, and $\operatorname{visible}$ tests whether the block's bounding sphere intersects the camera frustum. 
A block is loaded into GPU memory if it is temporally active and visible for at least one camera in $\mathcal{C}$.

The rendering and optimization speed of this scheme depends on blocks remaining compact in both space and time.
Existing out-of-core \ac{gs} methods use different forms of spatial organization and model growth.
TideGS~\cite{zhong2026tidegs} partitions Morton-ordered Gaussians into fixed spatial blocks and, in its large-scale setting, relies on a dense initialization without training-time densification or relocation.
CLM-GS~\cite{zhao2026clm} supports standard adaptive densification, while A LoD of Gaussians~\cite{windisch2026ALoDOfGaussians} grows and restructures a spatial Gaussian hierarchy through leaf spawning and respawning. 

However, all three methods organize Gaussian access using only spatial information and do not account for temporal support or motion.
Their direct extension to dynamic scenes may therefore load temporally inactive Gaussians and lose spatial locality as Gaussian positions evolve. 
This is particularly limiting in dynamic regions, where incomplete initialization and Gaussian motion reduce block compactness. 
Spatial-only grouping may place Gaussians with substantially different lifetimes in the same block, enlarging block temporal support and weakening temporal culling.

To address these limitations, \ac{doom-gs} updates both the Gaussian population and the block layout during training, with several operations.
Cloning increases the model capacity, relocation reuses inactive slots, and periodic repacking restores block compactness using 4D Morton ordering.
Cloning and relocation are performed during the densification phase and operate on the Gaussians in the current working set $\WorkingSet(\mathcal{C})$.
For Gaussian $i$, $\nabla_{\mean_i}\mathcal{L}$ denotes the gradient of the current batch loss with respect to its 3D mean $\mean_i$.
Each Gaussian is assigned the score
\begin{equation}
s_i
=
\frac{1}{2}
\left(
\frac{
\left\|\nabla_{\mean_i}\mathcal{L}\right\|_2
}{
\max_{j\in\WorkingSet(\mathcal{C})}
\left\|\nabla_{\mean_j}\mathcal{L}\right\|_2
}
+
\opacity_i
\right).
\end{equation}
Targets for both operations are sampled in proportion to $s_i$. 
This score favors Gaussians with large 3D position gradients and high opacity.

\compactparagraph{Cloning.}
Cloning grows the model toward $\NumberOfGaussians_{\max}$. 
Source Gaussians are sampled in proportion to $s_i$, and their opacity and scale are split between the source and its copies following 3DGS-MCMC~\cite{kheradmand20243d}. 
Each clone inherits the source's temporal and motion parameters. 
The clones are then sorted by their 4D Morton codes and appended to the store.

\compactparagraph{Relocation.}
Relocation reuses inactive Gaussian slots instead of pruning them, keeping the model size fixed. 
A Gaussian is considered inactive if its opacity falls below $0.005$, or if its opacity remains below $0.01$ and its 3D mean $\mean_i$ receives no position gradient for $2{,}000$ iterations.
Its slot is reassigned to a target sampled using $s_i$, provided that the opacity of every resulting copy remains at least $0.01$. 
To preserve spatial locality, each target reuses a dead Gaussian slot from the Morton cell with an available slot. 
The relocation is performed only if the dead Gaussian's position is within one median block diameter of the target.
At most $5\%$ of the working set is relocated per step.

\compactparagraph{Repacking.} 
Cloning and relocation do not fully preserve block compactness, as clones may originate from distant regions and repeated relocation can degrade spatial locality. We repacks the store every $2{,}000$ iterations by reordering all Gaussians and rewriting them into full blocks. 
A 4D Morton code over the normalized 3D means and temporal centers groups Gaussians that are close in space and time. 
To avoid mixing substantially different lifetimes, Gaussians active for at least half of the sequence are packed using spatial Morton ordering. 
The remaining Gaussians are divided into four duration classes with successively halved ranges and ordered within each class using a 4D Morton code with class-normalized temporal coordinates. 

\subsubsection{Camera Scheduling for Dynamic Gaussians}
One major factor of training speed is how often we can reuse blocks that are already resident in GPU memory. 
Since training is conducted on a series of data batches, consecutive batches with similar working sets require fewer new block transfers and are therefore processed more efficiently.
A dynamic capture contains $\NumberOfCams\times\NumberOfFrames$ views $(c,f)$, where $c$ is the camera and $f$ is the frame at time $t_f$. 
Random ordering provides little reuse between batches. 
CLM-GS~\cite{zhao2026clm} optimizes the order only within each batch, whereas TideGS~\cite{zhong2026tidegs} orders views by camera pose. 
For videos from stationary cameras, this repeatedly selects the same camera at different frames, reducing viewpoint diversity and harming optimization and image quality.

\ac{doom-gs} instead uses a global greedy scheduling strategy illustrated in Figure~\ref{fig:doomgs}.
We discretize the scene into a voxel grid and let $\mathcal{V}_c$ denote the occupied voxels intersecting the viewing frustum of camera $c$. 
The overlap between cameras $c$ and $c'$ is measured by the Jaccard similarity $J(c,c')$ of their voxel sets. 
Scheduling proceeds in passes, with each camera selected exactly once per pass. 
Given the current view $(c,f)$, we select the unused camera $c'$ that maximizes $J(c,c')$.
We then select its least-used remaining frame, breaking ties by choosing the timestamp closest to $t_f$. 
The resulting sequence is divided into batches. 
This schedule increases block reuse between consecutive batches, avoids repeated updates from the same camera, and distributes training views across the full sequence.

\subsubsection{Point Splatting with Gradient Estimators}
\label{sec:point_splatting_gradient_estimators}
\begin{figure}[b]
    \centering
    \includegraphics[width=\columnwidth]{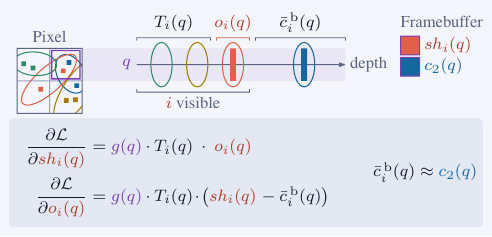}
    \caption{\textbf{Point-splatting gradients at subpixel $q$.}
    The depth expansion shows the transmittance $T_i(q)$, opacity $\opacity_i(q)$, and expected color $\bar{c}^{\,\mathrm b}_i(q)$ behind Gaussian $i$.
    The framebuffer stores the visible point and the nearest point from a different Gaussian, whose color $c_2(q)$ estimates the color behind $i$.}
    \label{fig:point_splatting_gradients}
\end{figure}

Even with block streaming, 3DGS rasterization requires all visible Gaussians to be in the GPU for sorting and alpha compositing, which often exceeds GPU memory already. 
We therefore employ the sort-free method \ac{gps}~\cite{rijsdijk2026gps}. However, GPS is designed only for forward rendering, therefore, we propose a gradient estimator for learning.
At subpixel $q$, GPS computes the color $I(q)$ using the nearest sampled point (see Section~\ref{sec:preliminaries}). This rendering process converges to the alpha compositing in expectation.
Indexing Gaussians front to back along a view ray, we partition this expectation around Gaussian $i$ as
\begin{equation}
\begin{gathered}
\mathbb{E}\!\left[I(q)\right]
=
\underbrace{
\sum_{j<i}
T_j(q)\opacity_j(q)\compcolor_j(q)
}_{\text{in front of }i}
\\
{}+
\underbrace{
T_i(q)\opacity_i(q)\compcolor_i(q)
}_{\text{Gaussian }i}
+
\underbrace{
\sum_{i<k}
T_k(q)\opacity_k(q)\compcolor_k(q)
}_{\text{behind }i}.
\end{gathered}
\label{eq:point_splatting_expected_color}
\end{equation}
The original GPS can be seen as a one-point Monte Carlo approximation of this expectation. The first term is not a function of $\opacity_i(q)$ and $\compcolor_i(q)$, while the second term is a function of both. The third term is a function of $\opacity_i(q)$:
\begin{equation}
T_k(q)
=
\left(1-\opacity_i(q)\right)
\underbrace{
T_i(q)
\prod_{i<n<k}
\left(1-\opacity_n(q)\right)
}_{\mathcal{B}_{ik}(q)}.
\label{eq:point_splatting_back_transmittance}
\end{equation}
The factor $\mathcal{B}_{ik}(q)$ is independent of $\opacity_i(q)$.
Substituting Equation~\ref{eq:point_splatting_back_transmittance} into the third term of Equation~\ref{eq:point_splatting_expected_color} allows the common factor $T_i(q)(1-\opacity_i(q))$ to be taken outside the sum.
The remaining term that will influence the gradient later defines the expected color behind Gaussian $i$ is
\begin{equation}
\bar{c}^{\,\mathrm b}_i(q)
=
\sum_{k>i}
\left[
\prod_{i<n<k}
\left(1-\opacity_n(q)\right)
\right]
\opacity_k(q)\compcolor_k(q).
\label{eq:point_splatting_background_color}
\end{equation}
Therefore, differentiating the third term in Equation~\ref{eq:point_splatting_expected_color} gives
\begin{equation}
\begin{aligned}
&\frac{\partial}{\partial\opacity_i(q)}
\left[
\sum_{k>i}
T_k(q)\opacity_k(q)\compcolor_k(q)
\right]
\\
&\quad=
\sum_{k>i}
-\mathcal{B}_{ik}(q)\opacity_k(q)\compcolor_k(q) = -T_i(q)\bar{c}^{\,\mathrm b}_i(q).
\end{aligned}
\label{eq:point_splatting_back_derivative}
\end{equation}
Combining this with the contribution of the second term in Equation~\ref{eq:point_splatting_expected_color}, and using the image-space gradient $g(q)=\partial\mathcal{L}/\partial I(q)$, gives
\begin{equation}
\begin{gathered}
\frac{\partial\mathcal{L}}
{\partial\compcolor_i(q)}
=
\opacity_i(q)T_i(q)g(q),
\\
\frac{\partial\mathcal{L}}
{\partial\opacity_i(q)}
=
T_i(q)g(q)
\left(
\compcolor_i(q)-\bar{c}^{\,\mathrm b}_i(q)
\right).
\end{gathered}
\label{eq:point_splatting_local_gradients}
\end{equation}
One challenge for the gradient estimator is estimating $\bar{c}^{\,\mathrm b}_i(q)$, which requires all the Gaussians behind Gaussian $i$ and is infeasible at scale. We propose a feasible approximation. The original 3DGS rasterizers obtain this quantity through ordered alpha compositing, whereas \ac{gps} retains no depth-sorted list and therefore cannot compute $\bar{c}^{\,\mathrm b}_i(q)$ directly.
To estimate this color, we use a framebuffer for each subpixel shown in Figure~\ref{fig:point_splatting_gradients}. The framebuffer stores the nearest and the second nearest point at each subpixel.
When Gaussian $i$ is visible, the second entry gives the color $c_2(q)$ that would be visible if points from $i$ were removed.
Independent point sampling gives $\mathbb{E}[c_2(q)\mid i\text{ is visible}]=\bar{c}^{\,\mathrm b}_i(q)$, so $c_2(q)$ is unbiased when $i$ is visible.

\begin{figure*}[t]
    \centering
    \includegraphics[width=\textwidth]{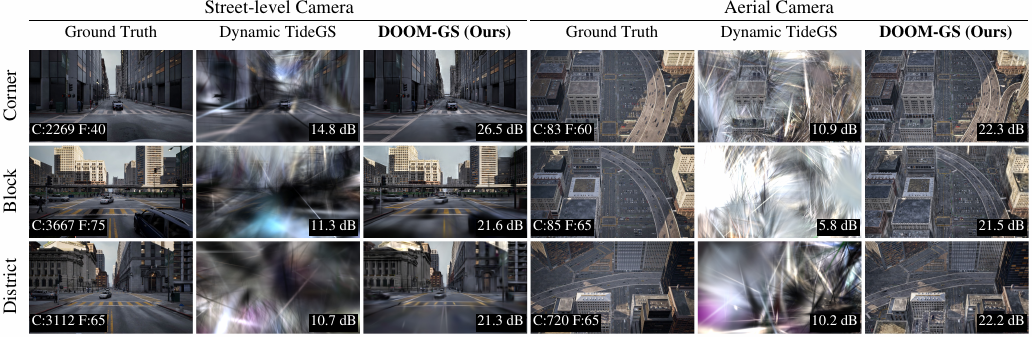}
    \caption{Qualitative comparison on held-out street-level and aerial views from \emph{Corner}, \emph{Block}, and \emph{District}. Dynamic TideGS exhibits streaking and haze, whereas \ac{doom-gs} better preserves scene structure and detail. C and F denote the camera and frame indices, respectively, and per-image PSNR are shown.}
    \label{fig:native_dynamic_experiments}
\end{figure*}

During the backward pass, we regenerate the same point samples and replay them against the completed framebuffer.
The depth test indicates whether Gaussian $i$ is visible in GPS, while $c_2(q)$ samples the color behind it.
The supplementary material provides the complete derivation and additional details.
Gradients of the remaining Gaussian parameters follow through the standard projection chain rule.

\compactparagraph{Streaming.} 
For each batch, we partition the working set into chunks that fit in GPU memory.
During the forward pass, the chunks are rendered sequentially into a separate framebuffer for each camera.
After all chunks are processed, we compute the rendered images, loss, and image-space gradients, retaining only the frame buffers and image gradients.
During the backward pass, each chunk is reloaded once, and its point samples are regenerated and replayed against all camera framebuffers.
The resulting gradients are accumulated across cameras, after which the Gaussian parameters are updated, written back to the block store, and released.
Since only one chunk and fixed-size per-camera buffers reside on the GPU, peak memory is independent of the number of Gaussians visible in a view.

\section{Experiments}
\label{sec:experiments}

\subsection{Experimental Setting}

\compactparagraph{Baselines.}
Existing out-of-core Gaussian-splatting methods primarily target static scenes. 
We therefore evaluate \ac{doom-gs} in two complementary comparison settings. 
First, we construct a dynamic variant of TideGS~\cite{zhong2026tidegs} by replacing its static primitives with the GraphiXS~\cite{yilmaz2026graphixs} dynamic representation while retaining its out-of-core framework, yielding a baseline with the same representation as ours. 
Second, we train CLM-GS~\cite{zhao2026clm}, TideGS~\cite{zhong2026tidegs}, and A LoD of Gaussians~\cite{windisch2026ALoDOfGaussians} independently for each frame. 
Together, these settings compare \ac{doom-gs} against both a dynamic extension of an existing method and per-frame static reconstruction approaches.

\compactparagraph{Dataset.}
MatrixCity~\cite{li2023matrixcity} is widely used for large-scale static reconstruction but provides no public dynamic scenes, so we extend its Small City map with procedurally generated pedestrians and vehicles while retaining the original camera poses.
The resulting \emph{DynamicMatrixCity} dataset contains three scenes of increasing scale, namely \emph{Corner} ($80\,m\times80\,m$), \emph{Block} ($260\,m\times150\,m$), and \emph{District} ($1{,}870\,m\times1{,}500\,m$).
Each camera records 120 frames at 30\,fps and $1920\times1080$ resolution.
We hold out 12 cameras in \emph{Corner} and 64 in each of \emph{Block} and \emph{District}, excluding all corresponding frames from training; further details are provided in the supplementary material.

\compactparagraph{Metrics.}
We report PSNR, SSIM, and LPIPS. 
We measure data reuse using the GPU hit rate, defined as the fraction of block accesses served by blocks already resident in GPU memory. 
For per-frame baselines, we additionally report flicker, which measures unwanted temporal changes in regions that remain static in the ground truth. 
Training efficiency is measured by wall-clock time excluding evaluation, summed over all frame models for per-frame baselines.

\begin{figure*}[t]
    \centering
    \includegraphics[width=\textwidth]{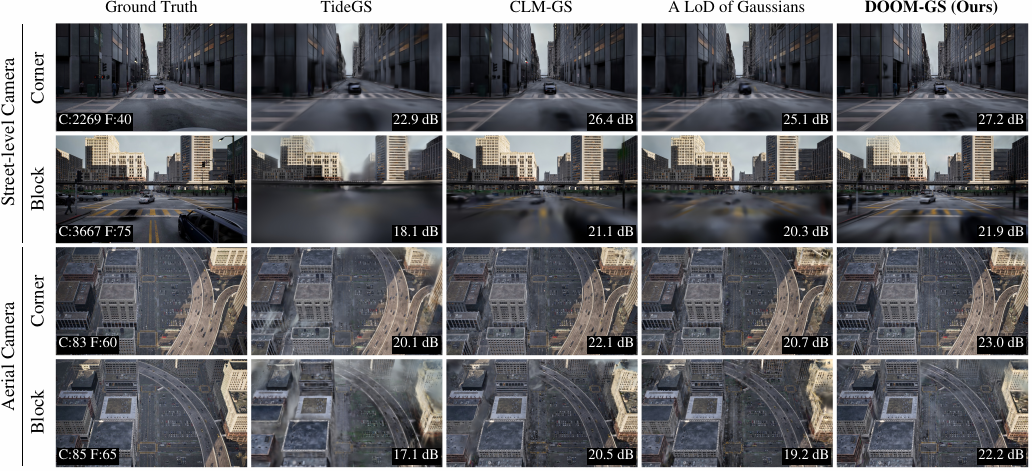}
    \caption{
    Qualitative comparison on held-out street-level and aerial views from \emph{Corner} and \emph{Block}.
    Each baseline uses a static model trained for the shown frame, whereas \ac{doom-gs} uses one dynamic model trained over the complete sequence. 
    C and F denote the camera and frame indices, respectively, and per-image PSNR are shown.
    }
    \label{fig:perframe_dynamic_experiments}
\end{figure*}

\subsection{Native Dynamic Comparison}
\label{sec:native_dynamic}

We compare \ac{doom-gs} with Dynamic TideGS on the three DynamicMatrixCity scenes.
Both methods optimize a single 20M-Gaussian model over the full 120-frame sequence, starting from the same initial point cloud and using the same dynamic Gaussian representation, learning rates, and densification schedule.
We train with batches of four images for 40K iterations on \emph{Corner} and \emph{Block}, and for 200K iterations on \emph{District}.
Both methods use a 1,800-block GPU budget, equivalent to 7.4M Gaussians, with peak GPU memory below 20\,GB.
Table~\ref{tab:native_dynamic} reports the quantitative results, and Figure~\ref{fig:native_dynamic_experiments} presents qualitative comparisons.
\begin{table}[b]
    \caption{
    Comparison on DynamicMatrixCity. Image-quality metrics are averaged over all frames of all held-out cameras. GPU hit rate is averaged over training. Memory is the peak GPU memory allocated during training in GB; time is training hours.
    }
    \label{tab:native_dynamic}
    \centering
    \small
    \setlength{\tabcolsep}{2.2pt}
    \resizebox{0.99\columnwidth}{!}{
    \begin{tabular}{@{}lcccccc@{}}
        \toprule
        Method
        & PSNR$\uparrow$
        & SSIM$\uparrow$
        & LPIPS$\downarrow$
        & Hit$\uparrow$
        & Mem.$\downarrow$
        & Time$\downarrow$ \\
        \midrule

        \multicolumn{7}{l}{\textit{Corner}} \\
        Dyn. TideGS
        & 15.60
        & 0.515
        & 0.570
        & 78\%
        & \textbf{18.0}
        & 29.5 \\
        \textbf{\ac{doom-gs} (Ours)}
        & \textbf{24.93}
        & \textbf{0.824}
        & \textbf{0.293}
        & \textbf{79\%}
        & 18.8
        & \textbf{15.9} \\

        \midrule
        \multicolumn{7}{l}{\textit{Block}} \\
        Dyn. TideGS
        & 9.28
        & 0.344
        & 0.687
        & 78\%
        & 18.9
        & 54.2 \\
        \textbf{\ac{doom-gs} (Ours)}
        & \textbf{23.19}
        & \textbf{0.721}
        & \textbf{0.447}
        & \textbf{89\%}
        & \textbf{16.8}
        & \textbf{37.2} \\

        \midrule
        \multicolumn{7}{l}{\textit{District}} \\
        Dyn. TideGS
        & 10.19
        & 0.286
        & 0.724
        & 75\%
        & 20.5
        & 317.4 \\
        \textbf{\ac{doom-gs} (Ours)}
        & \textbf{19.35}
        & \textbf{0.658}
        & \textbf{0.534}
        & \textbf{86\%}
        & \textbf{12.6}
        & \textbf{228.5} \\

        \bottomrule
    \end{tabular}
    }
\end{table}
\ac{doom-gs} substantially improves reconstruction quality across all three scenes.
It increases PSNR by 9.3\,dB on \emph{Corner}, 13.9\,dB on \emph{Block}, and 9.2\,dB on \emph{District}.
The improvements are consistent across the remaining metrics.
LPIPS decreases from 0.570 to 0.293 on \emph{Corner}, while SSIM increases from 0.344 to 0.721 on \emph{Block} and from 0.286 to 0.658 on \emph{District}.
The qualitative comparisons in Figure~\ref{fig:native_dynamic_experiments} similarly show cleaner street-level and aerial renderings.
\ac{doom-gs} also improves the GPU hit rate, reaching 89\% on \emph{Block} compared with 78\% for Dynamic TideGS.
It reduces training time by 46\% on \emph{Corner}, 31\% on \emph{Block}, and 28\% on \emph{District}.
Peak GPU memory is similar on \emph{Corner}, while \ac{doom-gs} reduces it from 18.9 to 16.8\,GB on \emph{Block} and from 20.5 to 12.6\,GB on \emph{District}.

The trajectory-based ordering of TideGS targets static multi-view captures and is not designed for stationary multi-camera videos.
Because all 120 frames from each camera share the same pose, Dynamic TideGS schedules them consecutively, reducing viewpoint diversity and visiting only 1,334 of 4,032 training cameras on \emph{Block} after 40K iterations.
In contrast, \ac{doom-gs} selects each camera once per pass, covering all training cameras every 1,008 iterations on \emph{Block} and 2,032 iterations on \emph{District}, thereby providing more balanced spatial and temporal supervision.

\begin{table}[b]
    \caption{
    Reconstruction quality and temporal consistency.
    Metrics are averaged over all frames of all held-out cameras.
    Flicker is reported in units of $10^{-3}$.
    }
    \label{tab:perframe_quality}
    \centering
    \small
    \setlength{\tabcolsep}{2.5pt}
    \begin{tabular}{@{}lcccc@{}}
        \toprule
        Method
        & PSNR$\uparrow$
        & SSIM$\uparrow$
        & LPIPS$\downarrow$
        & Flicker$\downarrow$ \\
        \midrule

        \multicolumn{5}{l}{\textit{Corner}} \\
        TideGS
        & 22.11
        & 0.744
        & 0.430
        & 29.18 \\
        CLM-GS
        & \textbf{25.70}
        & 0.830
        & 0.314
        & 25.73 \\
        A LoD of Gaussians
        & 24.04
        & 0.789
        & 0.384
        & 19.84 \\
        \textbf{\ac{doom-gs} (Ours)}
        & 25.33
        & \textbf{0.837}
        & \textbf{0.308}
        & \textbf{2.79} \\

        \midrule
        \multicolumn{5}{l}{\textit{Block}} \\
        TideGS
        & 19.50
        & 0.588
        & 0.613
        & 36.64 \\
        CLM-GS
        & 21.32
        & 0.670
        & 0.526
        & 50.96 \\
        A LoD of Gaussians
        & 21.69
        & 0.662
        & 0.547
        & 18.24 \\
        \textbf{\ac{doom-gs} (Ours)}
        & \textbf{23.02}
        & \textbf{0.712}
        & \textbf{0.477}
        & \textbf{2.89} \\

        \bottomrule
    \end{tabular}
\end{table}

Dynamic TideGS renders each view using only the visible blocks that fit within its 1,800-block GPU budget.
Coverage is initially 100\% because the complete model fits, but append-only densification and relocation reduce block locality as the model grows.
Near the end of training, a view contains around 4,800 of the 4,883 blocks, so only about 37\% of its visible blocks can reside on the GPU, yielding 43-48\% average coverage across the three scenes.
These incomplete renderings produce inconsistent gradients and the streaking and haze visible in Figure~\ref{fig:native_dynamic_experiments}. 
\ac{doom-gs} maintains 100\% coverage by streaming all visible blocks through point splatting, so the GPU budget bounds only simultaneous residency, not the number of primitives rendered per view. 
Its space-time block organization also preserves locality, reducing the visible set on \emph{Block} to approximately 2,000 blocks per view and increasing the GPU hit rate from 78\% to 89\%.

\begin{table}[t]
    \caption{
    Cost of reconstructing the complete 120-frame sequence.
    Per-frame methods store 120 independent models.
    Mem. is the peak GPU memory, time is the training time summed over
    all frames, in hours.
    }
    \label{tab:perframe_efficiency}
    \centering
    \small
    \setlength{\tabcolsep}{3.5pt}
    \begin{tabular}{@{}lccc@{}}
        \toprule
        Method
        & Total Gauss.$\downarrow$
        & Mem.$\downarrow$
        & Time$\downarrow$ \\
        \midrule

        \multicolumn{4}{l}{\textit{Corner}} \\
        TideGS
        & 300M
        & 2.6
        & 71.4 \\
        CLM-GS
        & 300M
        & \textbf{1.5}
        & \textbf{20.0} \\
        A LoD of Gaussians
        & 300M
        & \textbf{1.5}
        & 54.7 \\
        \textbf{\ac{doom-gs} (Ours)}
        & \textbf{2.5M}
        & 2.4
        & 22.9 \\

        \midrule
        \multicolumn{4}{l}{\textit{Block}} \\
        TideGS
        & 300M
        & 2.4
        & 76.0 \\
        CLM-GS
        & 300M
        & 1.6
        & 30.6 \\
        A LoD of Gaussians
        & 300M
        & \textbf{1.5}
        & 43.9 \\
        \textbf{\ac{doom-gs} (Ours)}
        & \textbf{2.5M}
        & 2.3
        & \textbf{14.5} \\

        \bottomrule
    \end{tabular}
\end{table}

\vspace{-0.2cm}
\subsection{Per-frame Dynamic Experiments}
\label{sec:perframe_dynamic}

We compare \ac{doom-gs} with per-frame optimized TideGS~\cite{zhong2026tidegs}, CLM-GS~\cite{zhao2026clm}, and A LoD of Gaussians~\cite{windisch2026ALoDOfGaussians} on \emph{Corner} and \emph{Block}. 
Each baseline treats the 120 frames as independent static scenes and trains one 2.5M-Gaussian model per frame for 5K iterations. 
We could not use \emph{District} as the required GPU memory for the baseline methods would be impractically large. 
\ac{doom-gs} instead trains one 2.5M-Gaussian dynamic model over the complete sequence for 40K iterations. 
All methods use batches of four images and at most 300 resident blocks.

Tables~\ref{tab:perframe_quality} and~\ref{tab:perframe_efficiency} report reconstruction quality and resource cost.
On \emph{Block}, \ac{doom-gs} achieves the best PSNR, SSIM, and LPIPS, improving PSNR by 1.3\,dB over the strongest per-frame baseline and 3.5\,dB over TideGS.
On \emph{Corner}, CLM-GS achieves 0.37\,dB higher PSNR, while \ac{doom-gs} obtains the best SSIM and LPIPS.
Compared with the most stable baseline, \ac{doom-gs} reduces flicker by $7.1\times$ on \emph{Corner} and $6.3\times$ on \emph{Block}.
Figure~\ref{fig:perframe_dynamic_experiments} shows qualitative trends for street-level and aerial views.

\ac{doom-gs} represents the complete sequence with one 2.5M-Gaussian model, compared with 120 per-frame models containing 300M Gaussians in total.
On \emph{Corner}, it is $3.1\times$ faster than TideGS and $2.4\times$ faster than A LoD of Gaussians, while CLM-GS is slightly faster at 20.0 hours compared with 22.9 hours.
On \emph{Block}, \ac{doom-gs} is the fastest method, outperforming TideGS, CLM-GS, and A LoD of Gaussians by $5.2\times$, $2.1\times$, and $3.0\times$, respectively.
Per-frame optimization duplicates static content and introduces frame-specific errors that appear as flicker.
By sharing parameters and supervision across time, \ac{doom-gs} uses $120\times$ fewer Gaussians and provides substantially greater temporal consistency, particularly on the larger \emph{Block} scene.
Although CLM-GS achieves slightly higher PSNR on \emph{Corner}, \ac{doom-gs} retains better structural, perceptual, and temporal consistency.

\subsection{Ablation Study}
\label{sec:ablation}
We conduct an ablation study by training 20M-Gaussian models on \emph{Corner} and \emph{Block} with all other settings fixed.
The variants disable space-time repacking, replace overlap-aware scheduling with random batches, and remove the locality constraint from relocation.
We report reconstruction quality, GPU hit rate, and training time.

\begin{table}[t]
    \caption{
    Ablation of \ac{doom-gs} at 20M Gaussians.
    GPU hit denotes the fraction of block accesses.
    Time is training hours.
    }
    \label{tab:ablation}
    \centering
    \small
    \setlength{\tabcolsep}{2.5pt}
    \resizebox{0.99\columnwidth}{!}{
    \begin{tabular}{@{}lccccc@{}}
        \toprule
        Variant
        & PSNR$\uparrow$
        & SSIM$\uparrow$
        & LPIPS$\downarrow$
        & Hit$\uparrow$
        & Time$\downarrow$ \\
        \midrule

        \multicolumn{6}{l}{\textit{Corner}} \\
        \ac{doom-gs}
        & 24.93
        & 0.824
        & 0.293
        & 79\%
        & 15.9 \\
        w/o space-time repacking
        & 25.21
        & 0.829
        & 0.290
        & 48\%
        & 107.0 \\
        w/o camera scheduling
        & 23.98
        & 0.774
        & 0.344
        & 52\%
        & 123.2 \\
        w/o locality relocation
        & 24.83
        & 0.826
        & 0.289
        & 55\%
        & 88.6 \\

        \midrule
        \multicolumn{6}{l}{\textit{Block}} \\
        \ac{doom-gs}
        & 23.19
        & 0.721
        & 0.447
        & 89\%
        & 37.2 \\
        w/o space-time repacking
        & 23.17
        & 0.720
        & 0.448
        & 51\%
        & 62.9 \\
        w/o camera scheduling
        & 22.82
        & 0.688
        & 0.467
        & 55\%
        & 76.7 \\
        w/o locality relocation
        & 23.23
        & 0.722
        & 0.447
        & 61\%
        & 53.8 \\

        \bottomrule
    \end{tabular}
    }
\end{table}

Space-time repacking and locality-constrained relocation improve efficiency while preserving reconstruction quality.
Removing them changes PSNR by at most 0.28 and 0.10\,dB, respectively, but lowers GPU hit rate by 31-38 and 24-28 percentage points and increases training time by $1.7$-$6.7\times$ and $1.4$-$5.6\times$.
Replacing the proposed schedule with random batches reduces PSNR by 0.95\,dB on \emph{Corner} and 0.37\,dB on \emph{Block}, lowers GPU hit rate by 27--34 percentage points, and increases training time by $2.1$-$7.7\times$.
These results show that repacking and local relocation preserve block locality, while camera scheduling additionally improves optimization.

\vspace{-0.2cm}
\section{Conclusion}
\label{sec:conclusion}
We introduced \ac{doom-gs}, the first out-of-core framework for large-scale dynamic Gaussian scenes. 
Our space-time-aware block organization, overlap-aware camera scheduling, and streaming point splatting with gradient estimators enable training under bounded GPU memory. 
Experiments on our new dataset \emph{DynamicMatrixCity} demonstrate improved reconstruction quality, temporal consistency, and training efficiency over dynamic and per-frame baselines.

\ac{doom-gs} requires an explicit motion representation and the full model to remain on SSD, limiting support for implicit deformation models and increasing storage demands at extreme scales.
A direct future direction is to use \ac{doom-gs} as an efficient, explicit, and persistent memory representation for world models~\cite{bruce2024genie,lu2025gaussianworldmodel,deng2026gaussiandwm}, potentially enabling training on moderate hardware.

{
    \small
    \bibliographystyle{ieeenat_fullname}
    \bibliography{main}
}

\end{document}